\documentclass[aps,a4paper,pra,oneside,
nobibnotes,nofootinbib,amsfonts,amssymb,amsmath,eqsecnum,showpacs,epsfig,11 pt]{revtex4}
\usepackage{graphicx}
\usepackage{placeins}
\usepackage{float}
\usepackage{epstopdf}
\usepackage{subfigure}
\usepackage{xcolor}
\usepackage[none]{hyphenat}
\usepackage{hyperref}

\newcommand{\sech}{\operatorname{sech}}

\begin{document}
\title{\textcolor{blue}{Nonlinear localized modes in $\mathcal{PT}$-symmetric Rosen-Morse potential well}\vspace{.2 cm}}
\author{{\bf Bikashkali Midya}}
\email{bikash.midya@gmail.com}
\author{{\bf Rajkumar Roychoudhury}}
\email{rroychoudhury123@gmail.com}
\affiliation{Physics \& Applied Mathematics Unit, Indian Statistical
 Institute, Kolkata 700108, India. \vspace{.75 cm}}
 \begin{abstract}
We report the existence and properties of localized modes described by nonlinear Schr\"odinger equation with complex $\mathcal{PT}$-symmetric
Rosen-Morse potential well.
Exact analytical expressions of the localized modes are found in both one dimensional and two-dimensional geometry with self-focusing and self-defocusing Kerr nonlinearity.
 Linear stability analysis reveals that these localized modes are unstable for all real
values of the potential parameters although corresponding linear Schr\"odinger eigenvalue problem possesses unbroken $\mathcal{PT}$-symmetry. This result has been verified by the direct numerical simulation of the governing equation.
The transverse power flow density associated with these localized modes has also been examined.
\end{abstract}

 \maketitle

\section{Introduction}
~~Recently, there has been considerable amount of attention to theoretical and experimental investigation of
light propagation in parity-time ($\mathcal{PT}$) symmetric optical media \cite{Ru+10,Ko10,Gu+09,Re+12,Ma08,Mu+08a,Be08,Lo09,Ka+10,Ga07,We+10,LCV11,Be+10,KGM08}.
The interest in study of such $\mathcal{PT}$-symmetric
optical materials has its roots in quantum mechanics: the paraxial equation of diffraction is mathematically equivalent to that of quantum
Schr\"odinger equation. Quantum mechanics requires that the spectrum of every physical observable should be real,
which of course are satisfied by Hermitian operators. However,
Bender and Boettcher \cite{BB98} pointed out that some non-Hermitian Hamiltonians with $\mathcal{PT}$-symmetry can also exhibit an entirely
real spectrum and may constitute unitary quantum systems without violating any of the axioms of quantum mechanics. Moreover, it has been shown that for a $\mathcal{PT}$-symmetric complex Hamiltonian, there may exist a threshold above which its eigenvalues are not
real but become complex, and the system undergoes a phase transition because of spontaneous $\mathcal{PT}$-symmetry breaking. In general the action of the parity $\widehat{P}$ and time $\widehat{T}$ operators is defined as $\hat{p} \rightarrow - \hat{p}$, $\hat{x} \rightarrow -\hat{x}$ and $\hat{p} \rightarrow - \hat{p}$, $\hat{x} \rightarrow \hat{x}$, $i \rightarrow - i$, respectively. A Hamiltonian
with a complex $\mathcal{PT}$-symmetric potential requires that the real part of the potential must be even function of position and the
imaginary part should be odd \cite{Be07}. In optics such complex $\mathcal{PT}$-symmetric structure can be designed
through a judicious designs that involve both optical gain/loss regions and the process of index guiding \cite{Gu+09,Ma08,Mu+08a}. In such settings, complex refractive index distribution
$n(x) = n_{R}(x) + i~ n_{I}(x)$ plays the role of an optical potential so that the index guiding $n_R(x)$
and the gain/loss profile $n_I(x)$ satisfy $n_R(x) = n_R(-x)$ and $n_I(x) = - n_I(-x)$, respectively.  Unusual exotic
phenomena like $\mathcal{PT}$ phase transition, band merging,
double refraction, non-reciprocity \cite{Ma08,Ma+10,MRR10}, and unidirectional invisibility \cite{Li+11,Mo13,Lo11} etc have been reported to exist in linear $\mathcal{PT}$-symmetric complex optical media.
Spontaneous $\mathcal{PT}$-symmetry
breaking has been experimentally observed
in active or passive $\mathcal{PT}$ dimers \cite{Gu+09,Ru+10} and periodic lattices \cite{Re+12}.
These findings, in turn, have stimulated considerable
research activity in the non-linear $\mathcal{PT}$-symmetric systems as well.

~~ In the nonlinear domain, a novel class of one and two dimensional localized modes were found to exist below and above the
phase transition point \cite{Mu+08b} and the interplay between the Kerr nonlinearity and the $\mathcal{PT}$ threshold was investigated \cite{Mu+08a}.
Subsequently, nonlinear modes are
studied in complex $\mathcal{PT}$-symmetric periodic \cite{Fa+12,NGY12}, Gaussian \cite{Hu+11}, Bessel \cite{HH12}, Scarf-II \cite{KMB12}, and
harmonic \cite{ZK12}
potentials, as well as in a harmonic trap with a
rapidly decaying $\mathcal{PT}$-symmetric imaginary component \cite{Ac+12}. Stable localized modes in a $\mathcal{PT}$-symmetric slab
waveguide with distributed gain and loss are found in \cite{TTA12}. Existence of optical solitons in
$\mathcal{PT}$-symmetric nonlinear couplers with gain/loss \cite{Al+12,DM11}, gap solitons in $\mathcal{PT}$-symmetric optical
lattices \cite{Zh+11} and
optical defect modes in $\mathcal{PT}$-symmetric potentials \cite{WW11} are also reported.  Stable 1D
and 2D bright spatial solitons are found to exist in defocusing Kerr media with $\mathcal{PT}$-symmetric Scarf II
potentials \cite{Sh+11}. Also, it has been found that the gray solitons
in $\mathcal{PT}$-symmetric potentials can be stable \cite{Li+11d}. However the existence of nonlinear localized modes in yet another important potential e.g.
complex $\mathcal{PT}$-symmetric Rosen-Morse well has not been reported so far. The complex Rosen-Morse potential well
is characterized by the same real component as the complex Scarf-II potential, however, its imaginary
component is different. In fact, in contrast with the real component, the imaginary potential component doesn't
vanish asymptotically, rather it tends to a finite value. This is the reason why the phenomenon of spontaneous breakdown of $\mathcal{PT}$-symmetry
is elusive in such system \cite{Zn00,LM09}. Nevertheless, the bound state energy eigenvalues of $\mathcal{PT}$-symmetric Rosen-Morse potential well undergoes
a shift from negative to positive domain for
certain range of parameters which controls the strength of the potential.

~~In this paper, we investigate the propagation of nonlinear beam in a single $\mathcal{PT}$ waveguide cell which is
characterized by the nonlinear Schr\"{o}dinger equation with complex Rosen-Morse potential well. Specifically, the existence of the spatial
localized modes have been reported in both one-dimensional and two-dimensional settings with self-focusing and self-defocusing Kerr nonlinearity.
 We have shown, with the help of linear stability analysis of the one-dimensional localized modes, that though the spontaneous breakdown of
$\mathcal{PT}$-symmetry
does not occur in complex Rosen-Morse well the localized modes corresponding to nonlinear Schr\"odinger equation are always unstable.
This linear instability has been verified by direct numerical
simulation of the governing
equation. The transverse power flow density associated with these nonlinear localized modes has also been examined.

\section{Localized Modes in $\mathcal{PT}$-symmetric complex Rosen-Morse well}
\subsection{Mathematical Model}
 We consider optical wave propagation in a Kerr nonlinear $\mathcal{PT}$-symmetric potential. In this case,
 $(1+1)$-dimensional optical beam propagation along longitudinal $z$ direction
 is governed by the following non-linear Schr\"odinger like equation \cite{Ga07,Mu+08a}
 \begin{equation}
i \frac{\partial \Psi}{\partial z} + \frac{\partial^2 \Psi}{\partial x^2} + \left[ V(x)+ i W(x)\right] \Psi + \sigma |\Psi|^2 \Psi =0. \label{e1}
\end{equation}
Here $\Psi(x,z)$ is slowly varying complex electric field envelop, $x$ is the transverse co-ordinate, and $\sigma=\pm 1$ represent the self-focusing and self-defocusing
nonlinearity respectively. $V(x)$ and $W(x)$ are the real and imaginary parts of the complex $\mathcal{PT}$-symmetric potential such that $V(-x) = V(x)$ and $W(-x) = -W(x)$.
Physically, $V(x)$ is responsible for
the bending and slowing down of light, and
$W(x)$ can lead to either amplification
(gain) or absorption (loss) of light within an optical
material.

The optical beam propagation in a single $\mathcal{PT}$ cell is important to understand light self-trapping in complex optical lattices.
In order to investigate the optical beam propagation in a single $\mathcal{PT}$ cell, we consider the
complex $\mathcal{PT}$-symmetric Rosen-Morse potential well as
\begin{equation}
\begin{array}{ll}
V(x) = -a (a + 1) \sech^2 x,\\
W(x) =  2 b \tanh x,
\end{array}\label{e2}
\end{equation}
where $a$ and $b$ characterizes the strength of the real and imaginary parts of the potential, respectively.
Both the real and imaginary part of this potential are shown in figure \ref{f1}a for the potential parameters $a=.75$ and $b=.8$. The linear
Schr\"odinger eigenvalue problem for the potential $\ref{e2}$ has been thoroughly studied in \cite{Zn00,LM09}.
It has been shown that all the bound state energy eigenvalues corresponding to the linear Schr\"odinger equation of complex Rosen-Morse well
are real so that the spontaneous breakdown
of $\mathcal{PT}$-symmetry never occur.
However, the energy eigenvalues
\begin{equation}
 \lambda_n = -(a-n)^2 + \frac{b^2}{(a-n)^2}, ~ n=0,1,... n_{max} < a.
\end{equation}
begin to shift from negative to the positive domain when the strength of the non-Hermiticity is increased. In fact
all the energy eigenvalues become positive whenever $\sqrt{|b|} > a^2$. Here, we search for stationary
solution of the nonlinear equation (\ref{e1}) in the form $\Psi(x,z) = \phi(x) e^{i \lambda z}$,
where $\lambda$ is the real propagation constant, and the complex function $\phi(x)$ satisfies following equation
\begin{equation}
\frac{d^2 \phi}{d x^2} - \left[a(a+1) \sech^2 x - 2 i b \tanh x \right] \phi + \sigma |\phi|^2 \phi = \lambda \phi. \label{e3}
\end{equation}
In the following we report the existence and linear stability of the localized modes of the above nonlinear equation (\ref{e3}) for both
the self-focusing and self-defocusing cases.

\subsection{Analytical solutions and their linear stability}
\subsubsection{Self-focusing case ($\sigma=1$)}
\begin{figure}[]
\centering
\includegraphics[width=4.25 cm,height=4.5 cm]{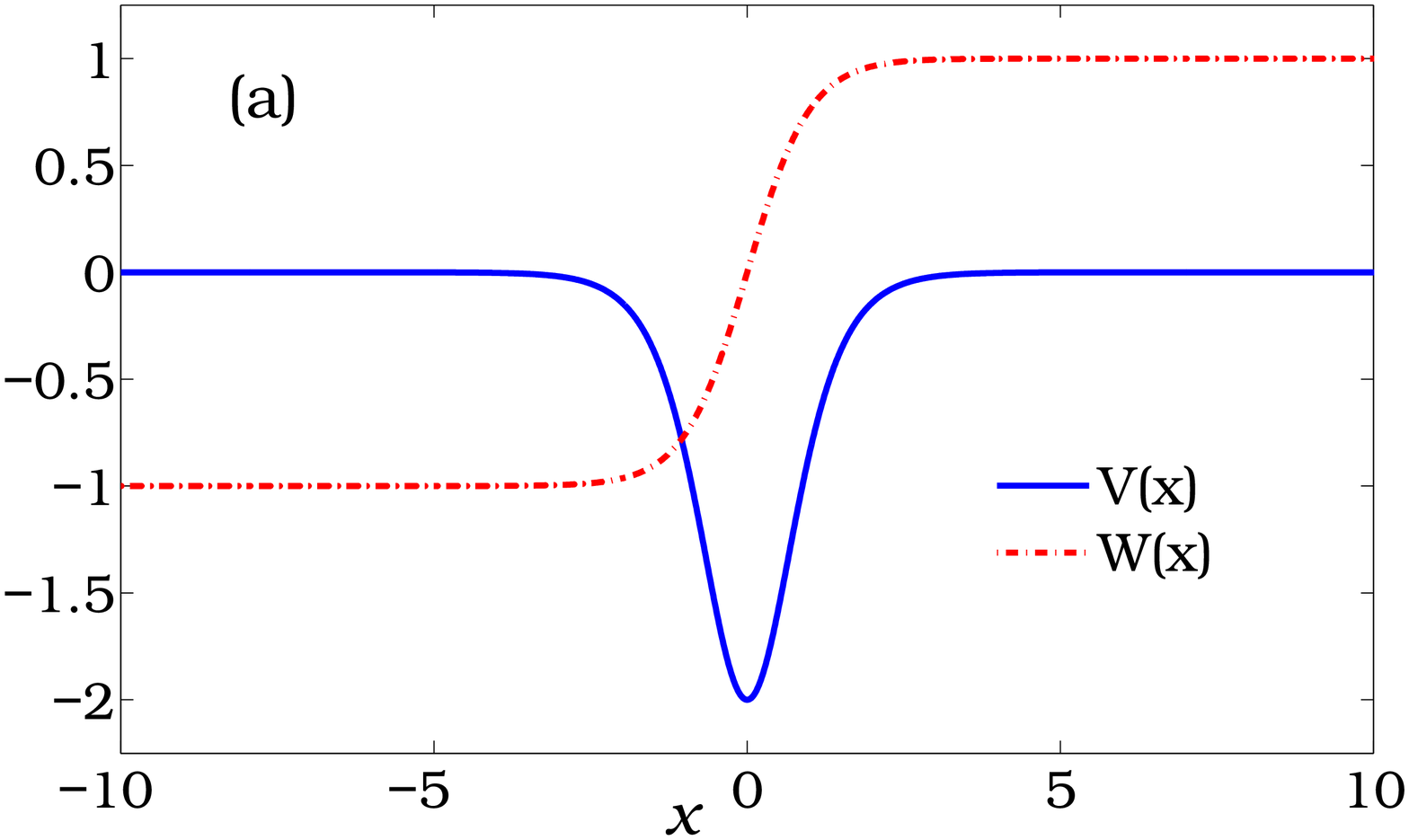} \includegraphics[width=4.3 cm, height=4.5 cm]{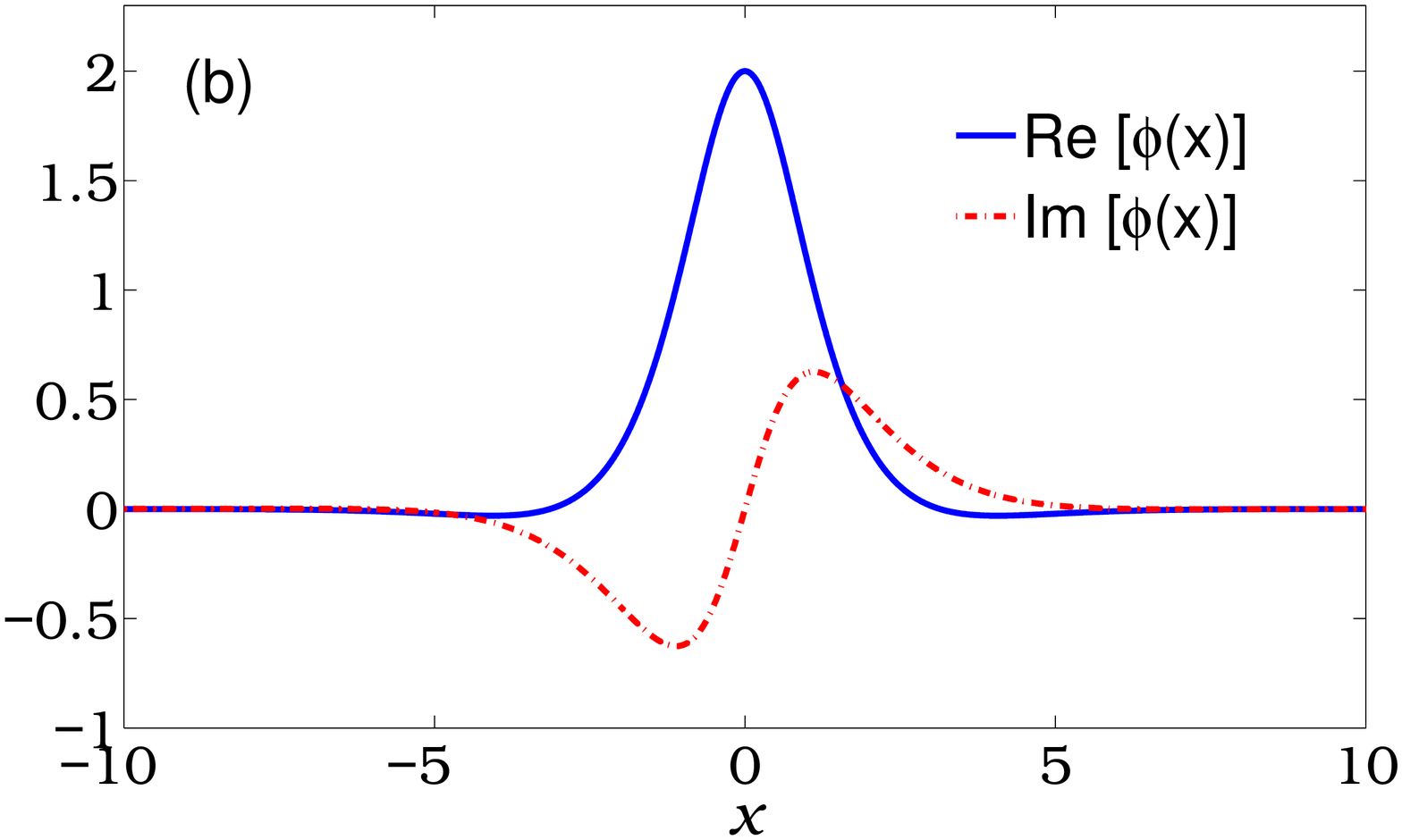}
 \includegraphics[width=4.25 cm,height=4.75 cm]{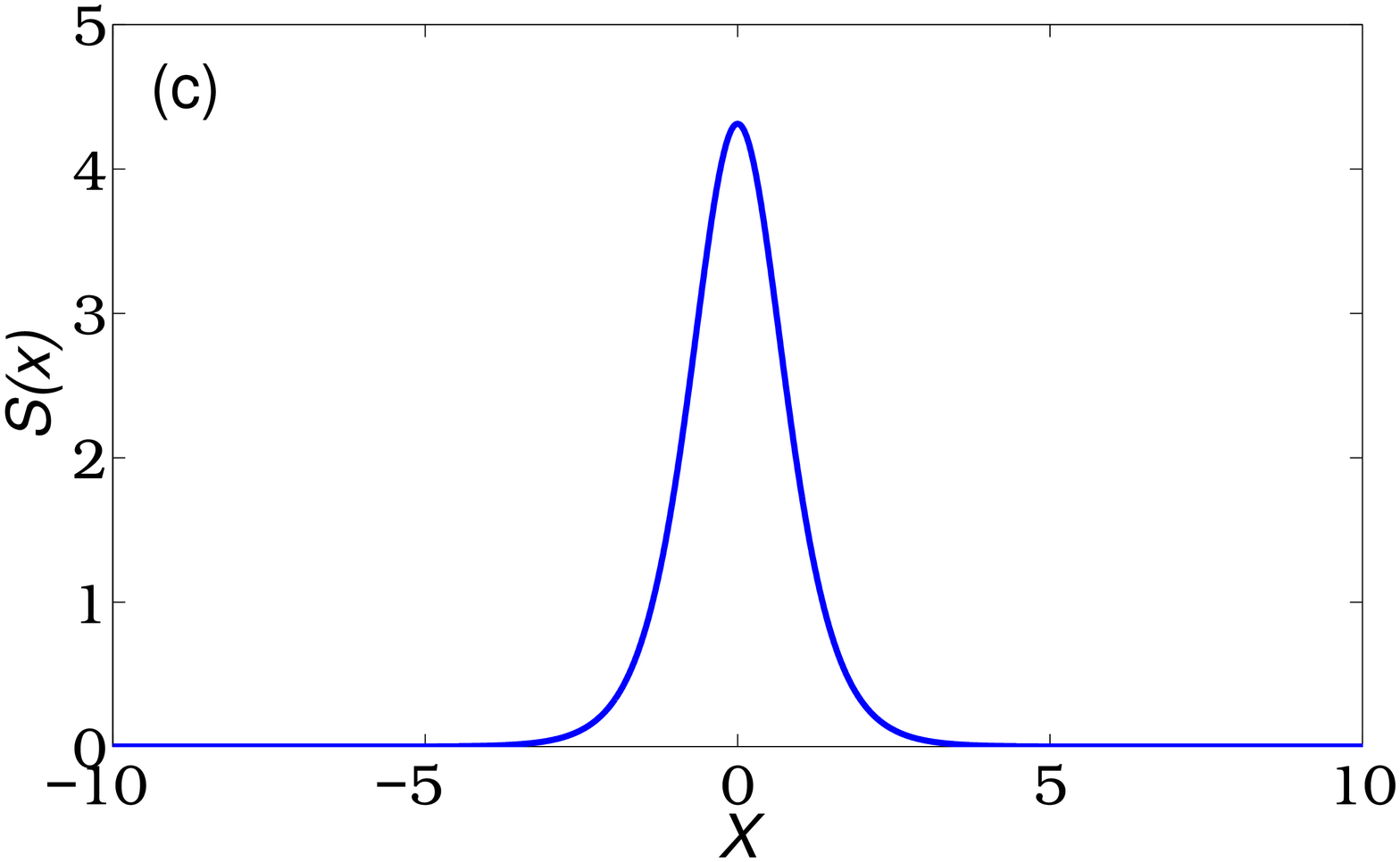}  \includegraphics[width=4.25 cm, height=4.5 cm]{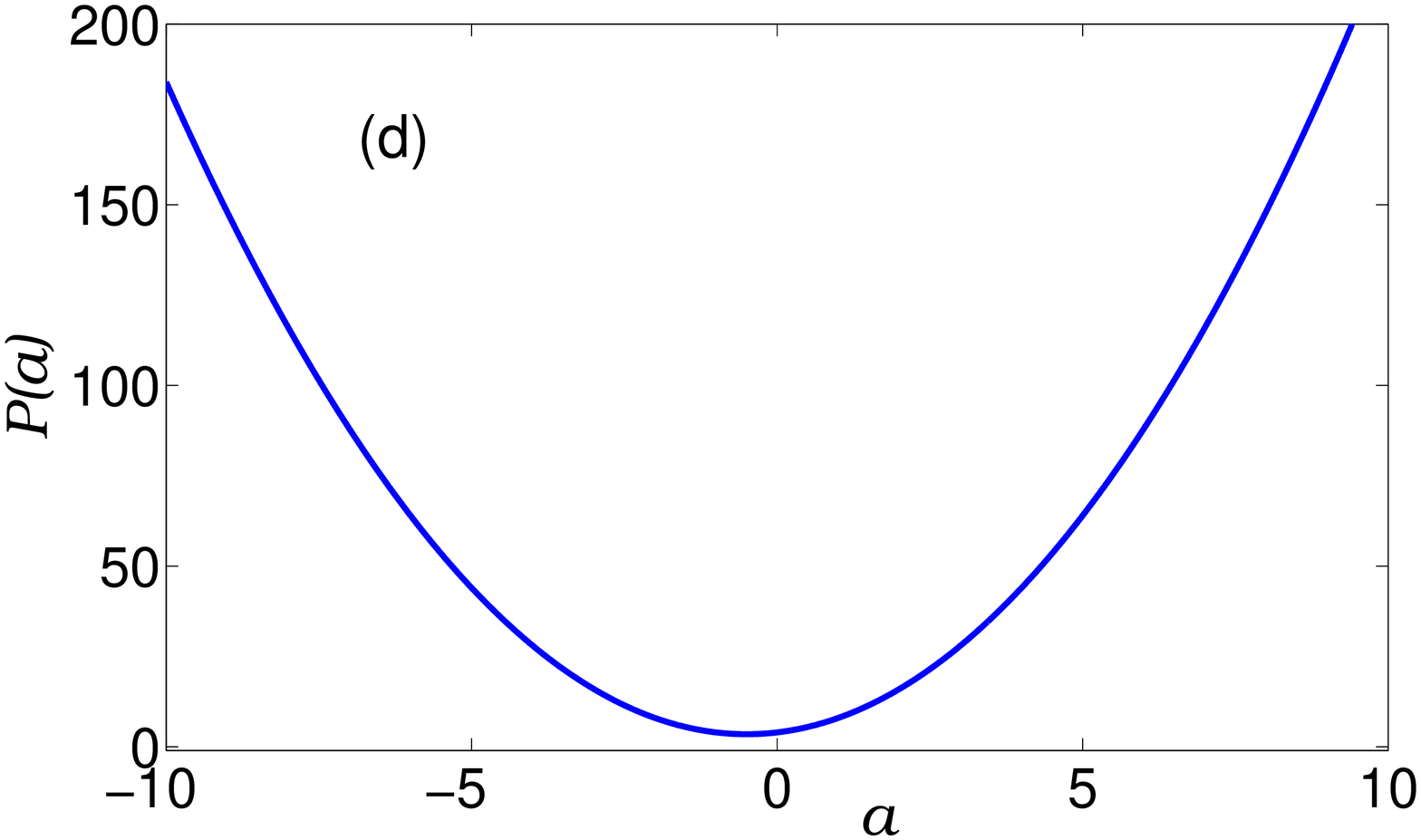}
\caption{(Color online) (a) Real and Imaginary parts of the Rosen-Morse Potential Well; (b) Real and Imaginary parts of the
localized modes $\phi(x)$ in the self-focusing medium; (c) The transverse power (Poynting vector) $S(x)$;  (d) Power $P(a)$ as a
function of potential parameter $a$. In (a), (b) and (c) we have considered $a=.75$,
$b=.8, \lambda =.36$ and $\sigma=1$.}\label{f1}
\end{figure}

For $\sigma=1$, equation (\ref{e3}) is found to admit an exact analytical expression of the localized mode of the form
\begin{equation}
\phi(x) = \sqrt{a^2+a+2}~ \sech x  ~e^{i b x}, \label{e4}
\end{equation}
where $\lambda = 1 - b^2.$ In figure \ref{f1}b, the real and imaginary parts of this spatial soliton have been shown
for $a= .75 $ and $b= .8$ and $\lambda = .36$. To focus on the properties of this non-linear solution, we examine following three quantities:
the transverse power flow density (Poynting vector) $S$
across the beam, the power $P$, and the linear stability of these localized modes. For the nonlinear modes given in equation (\ref{e4}), the Poynting vector
$S = \frac{i}{2} (\phi \phi_x^* -\phi^* \phi_x) = b (a^2+a+2) \sech^2 x$ depends on the sign
of the strength, $b$, of the imaginary part of the potential. It may be both negative and positive for negative and positive values of $b$ respectively.
However, we consider only positive values  of $b$ in which case $S$ is everywhere positive
and the power flow in the $\mathcal{PT}$ cell is in one direction, i.e. from the gain towards loss domain. For $a=.75$, $b=.8$, the transverse power
flow is shown in figure
\ref{f1}(c). For the localized modes (\ref{e4})
the power $P$ is calculated as
\begin{equation}
 P(a,b) = \int_{-\infty}^{\infty} |\phi(x)|^2 dx = 2 (a^2+a+2).
\end{equation}
Clearly, the power is independent of the parameter $b$. In figure \ref{f1}(d) we have shown
the power $P$ as a function of the real potential strength parameter $a$. It remains positive for all values of the parameter $a$. The Power
increases with the increasing absolute value of the
parameter $a$ and becomes minimum for $a=-.5$. At $a=-.5$, the
amplitude of the localized modes is also minimum.
\FloatBarrier

\begin{figure}[ht!]
\includegraphics[width=4.25 cm,height=4.5 cm]{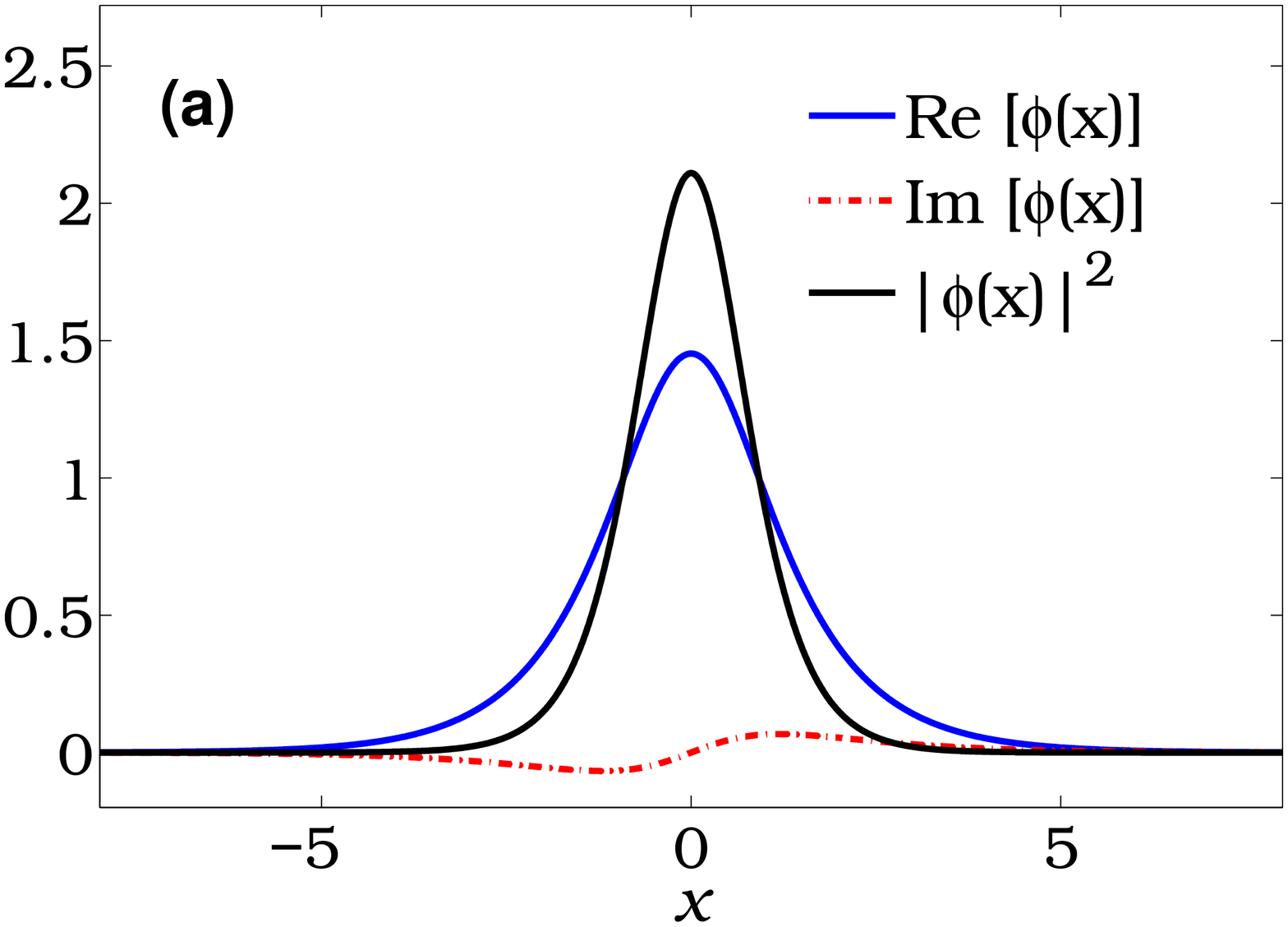}  ~~~~~~ \includegraphics[width=5.5cm,height=5.5cm]{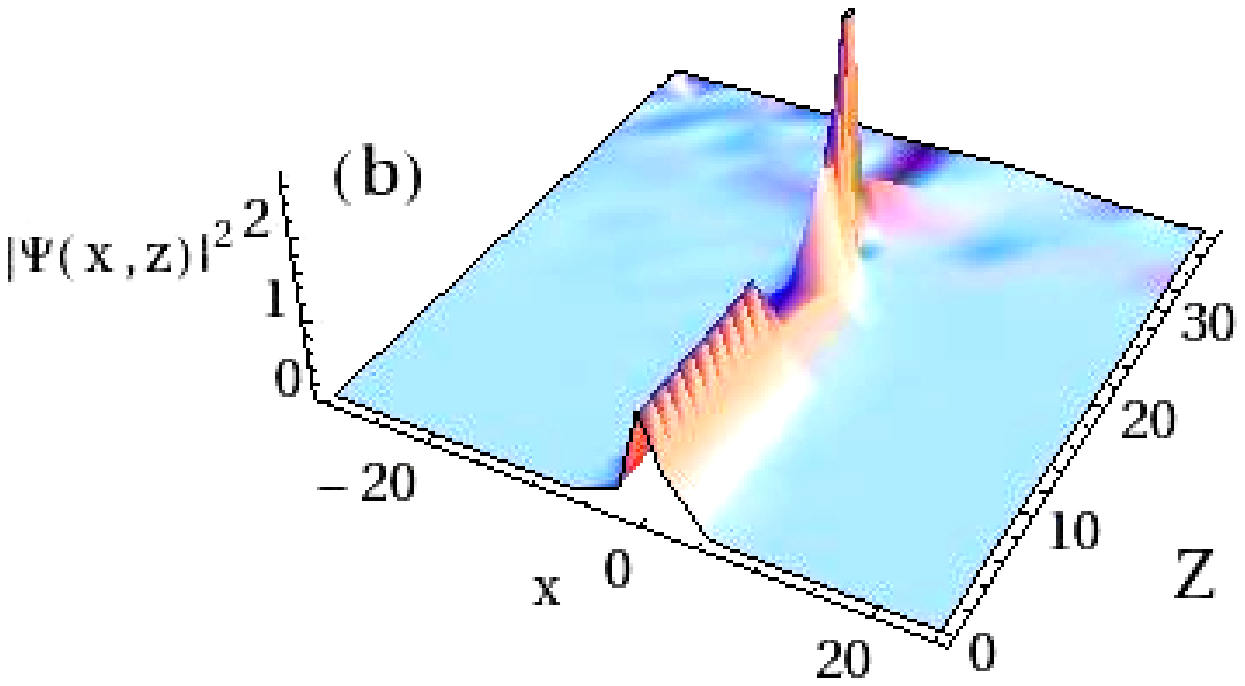}
~~~~~~~~\includegraphics[width=4 cm, height=4.5 cm]{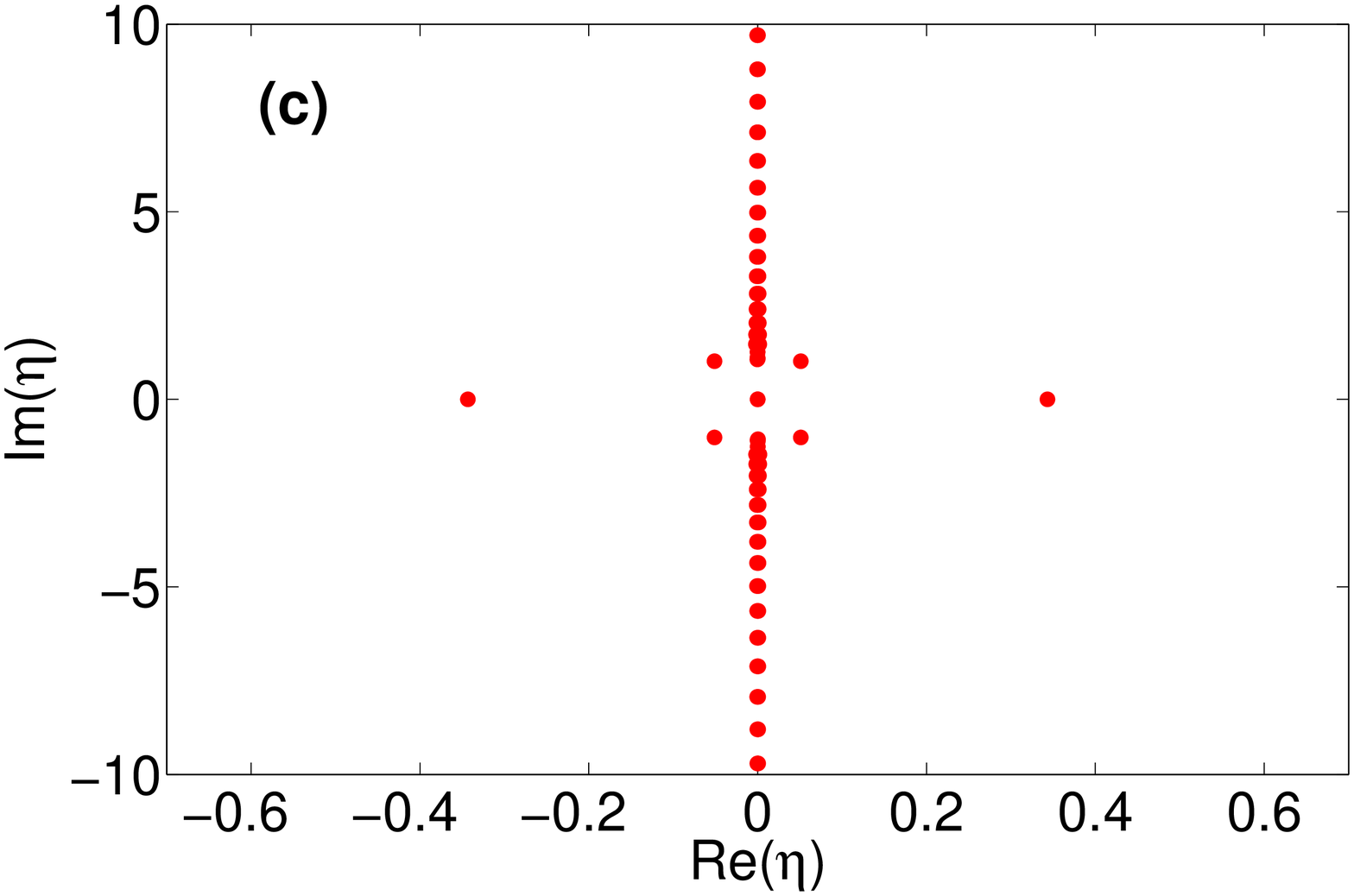} ~
\includegraphics[width=4.25 cm,height=4.5 cm]{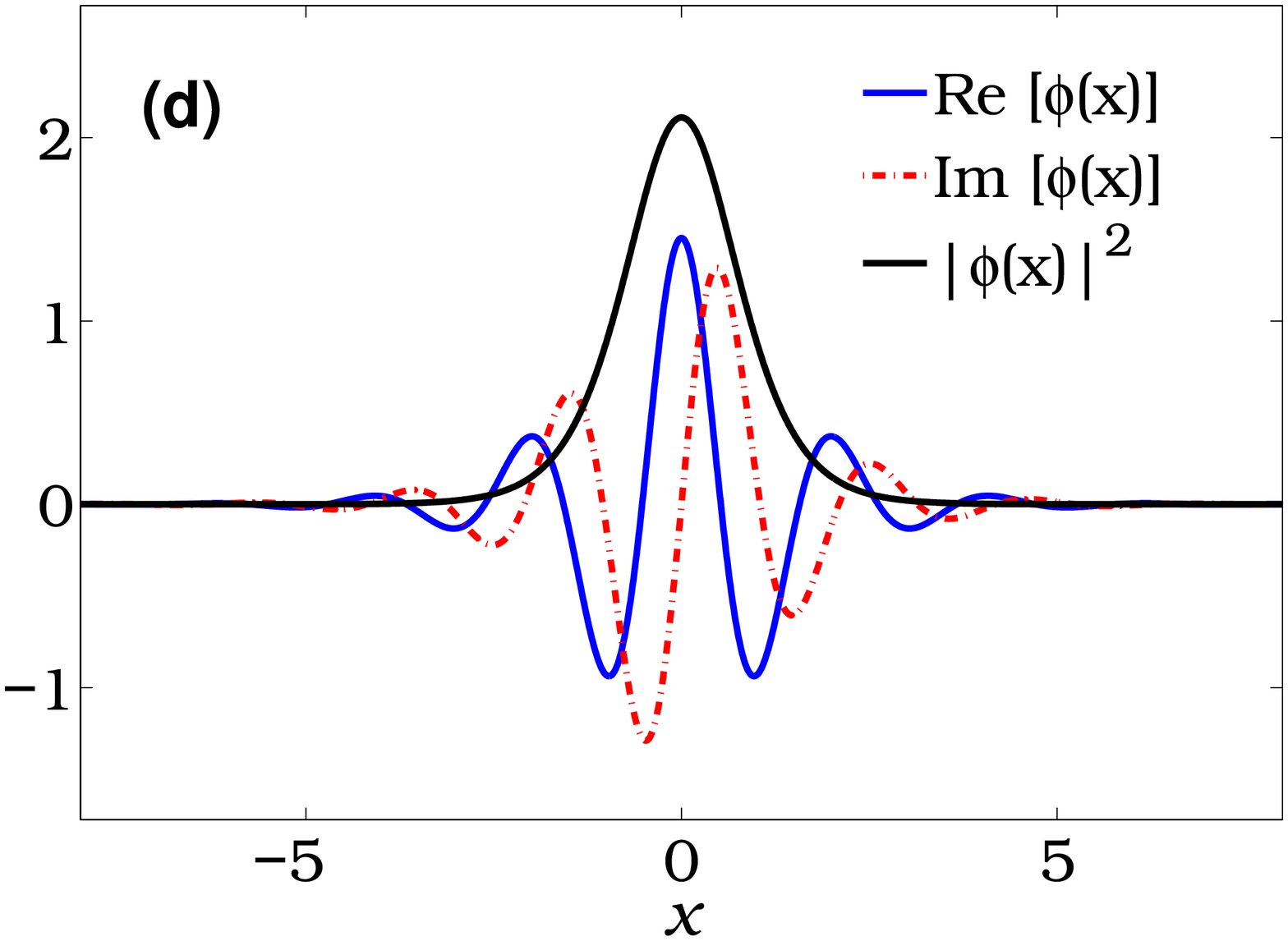}  ~~~~~~~ \includegraphics[width=5.5cm,height=5.5cm]{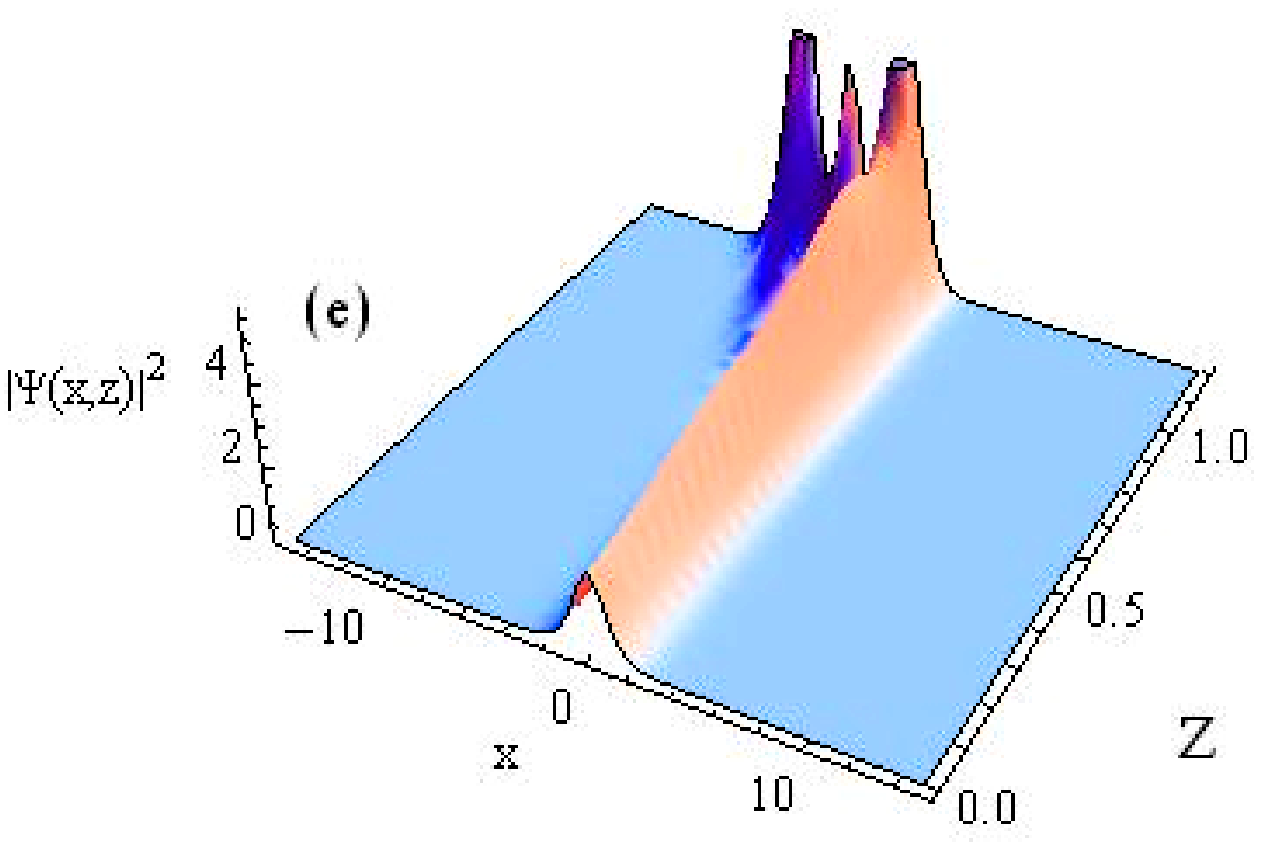} ~~~~~~~~~~\includegraphics[width=4 cm, height=4.5 cm]{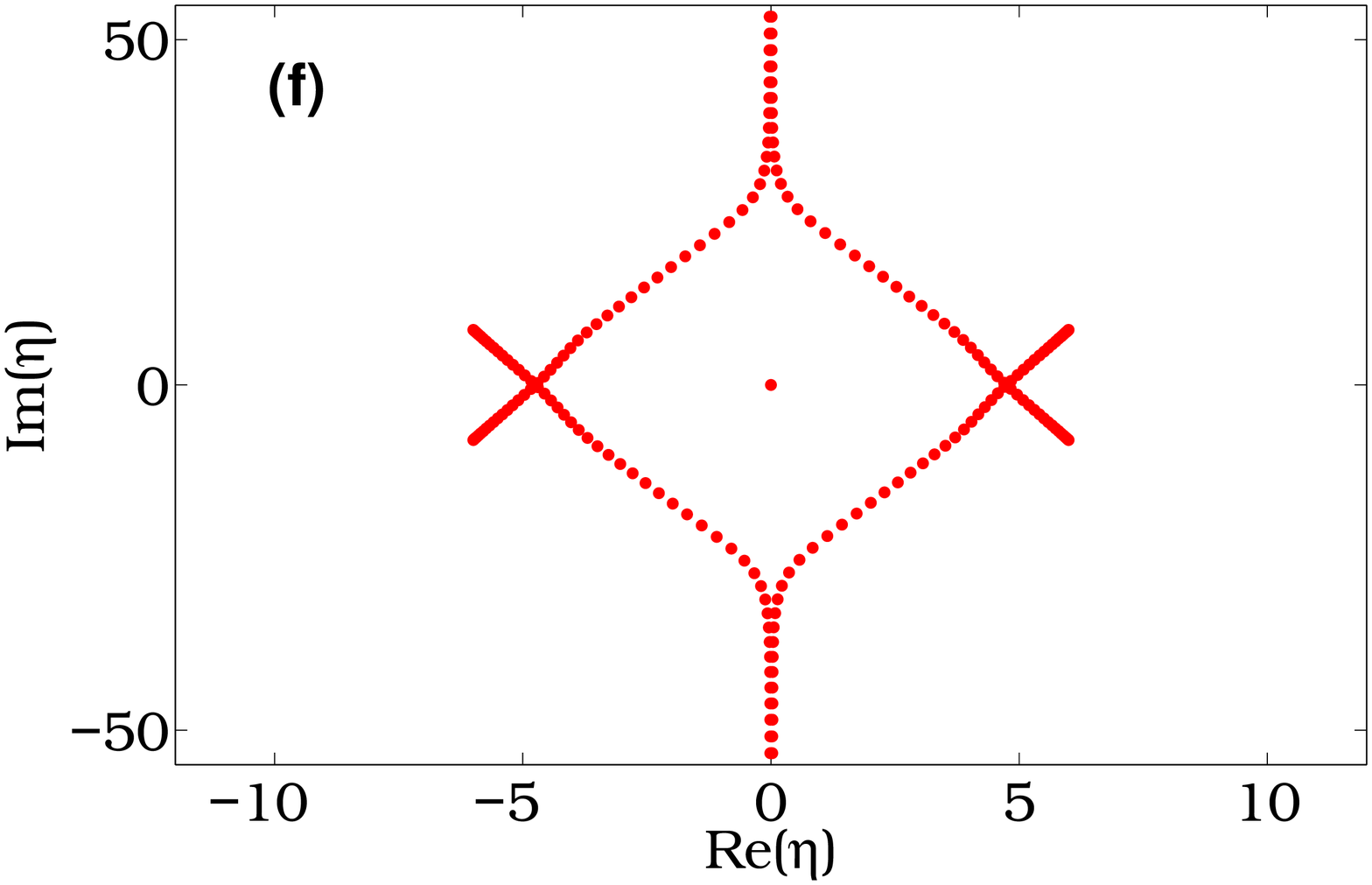}
\caption{(Color online) (a) Plots of the real (solid blue curve) and imaginary (dotted-dashed red curve)
parts of $\phi(x)$ and $|\phi(x)|^2$ (solid black curve) in self focusing medium;
(b) The evolution of field intensity $|\Psi(x,z)|^2$ (c) Numerically computed stability spectra corresponding to the figure \ref{f2}a. In all these cases we have considered $a=.1, b=.03, \lambda =.999$ and $\sigma=1$.  (d), (e) and (f) Plots of the
same quantities as in figure \ref{f2}(a), (b) and (c) respectively, for
$a=.1, b=3, \lambda=-8, \sigma=1.$}\label{f2}
\end{figure}

 In order to determine the linear stability properties of the self-trapped localized modes obtained here,
we consider small perturbation to the solution $\Psi(x,z)$, of the form \cite{KMB12,NGY12,ZKK11}
\begin{equation}
 \Psi(x,z) = \phi(x) e^{i \lambda z} + \left\{[ f(x) + g(x) ] ~e^{\eta z} + [f^*(x) -g^*(x)]~ e^{\eta^* z}\right\}  e^{i \lambda z}\label{e30}
\end{equation}
where $f(x)$ and $g(x)$ are infinitesimal perturbation eigen-functions such that $|f|,|g| \ll |\phi|$,
$\eta$ stands for the perturbation growth rate. By linearizing the equation (\ref{e1}) around the
localized solution $\phi(x)$, we find that the functions $f$ and $g$ satisfy the following eigenvalue problem
\begin{equation}
 \left( \begin{array}{cc}
0 & ~\hat{\mathcal{L}}_1  \\
 \hat{\mathcal{L}}_2 & ~0 \\
\end{array} \right)   ~~~ \left( \begin{array}{c}
f   \\
g \\
 \end{array} \right)= - i \eta
 \left( \begin{array}{c}
f \\
g \\
 \end{array} \right)\label{e12}
\end{equation}
where $\hat{\mathcal{L}}_1 = \partial_{xx} + (V+i W) + \sigma |\phi|^2 - \lambda$ and
$\hat{\mathcal{L}}_2 = \partial_{xx} + (V+i W) + 3 \sigma |\phi|^2 - \lambda$. The linear stability of the localized modes $\phi(x)$
depends on the nature of the eigenvalue $\eta$. The $\mathcal{PT}$-symmetric nonlinear localized mode is unstable if $\eta$ has any positive real part, because for
$\Re(\eta) > 0$ the corresponding
perturbed nonlinear eigenmodes (\ref{e30}) would grow exponentially with $z$. The eigenvalues $\eta$ can be obtained by solving equation
(\ref{e12}) with the help of several
numerically techniques \cite{Ya08}. In this paper we have used Fourier collocation method \cite{Ya10}. Our numerical investigations corresponding
to the nature of the eigenvalue $\eta$ reveal that $\eta$ never becomes purely imaginary. It has always non vanishing positive
real part for all real values of
potential parameters  $a$ and $b$.
This implies that the nonlinear localized modes obtained here
are always unstable. The results of linear stability analysis are corroborated
by direct numerical simulations of Eq. (\ref{e1}) using the solution (\ref{e4}) as initial condition i.e. $\Psi(x,0) = \phi(x)$. In figure \ref{f2} we have shown
the localized modes $\phi(x)$, unstable intensity profiles $|\Psi(x,z)|^2$ and corresponding linear stability spectra for very small and large values of the parameter $b$.
As expected, localized modes which are
predicted to be unstable fail to maintain their original
shapes. The reason behind such instability of the localized modes is that unlike the real part, the imaginary part of the
Rosen-Morse potential well does not vanish asymptotically. Therefore gain/loss remains in the system even far from the place of localization and
any small fluctuations of the field is amplified/absorbed, eventually
leading to instability.
\FloatBarrier
It is worth mentioning here that unlike the $\mathcal{PT}$-symmetric Scarf II potential
(for which the solitons are stable below the certain critical value of the imaginary
potential component
and become unstable above this critical value \cite{Mu+08a,KMB12}), localized modes in the complex Rosen-Morse potential well, discussed here, are
linearly unstable for all real
values of non-Hermiticity parameter $b$. This result is valid in spite of the fact that the Hamiltonian corresponding to the
linearized version of equation (\ref{e3}) possesses unbroken
$\mathcal{PT}$-symmetry (all energy eigenvalues are real). Nevertheless, the parameters range $\sqrt{|b|} >a^2$,
for which the transition from
the negative to positive energy corresponding to the linear Schr\"odinger eigenvalue problem takes place, does not affect the instability of the localized modes.
Only the magnitude of the localized modes differs.

\subsubsection{Self-defocusing case ($\sigma=-1$)}
\begin{figure}[h]
\includegraphics[width=4.5 cm,height=4.5 cm]{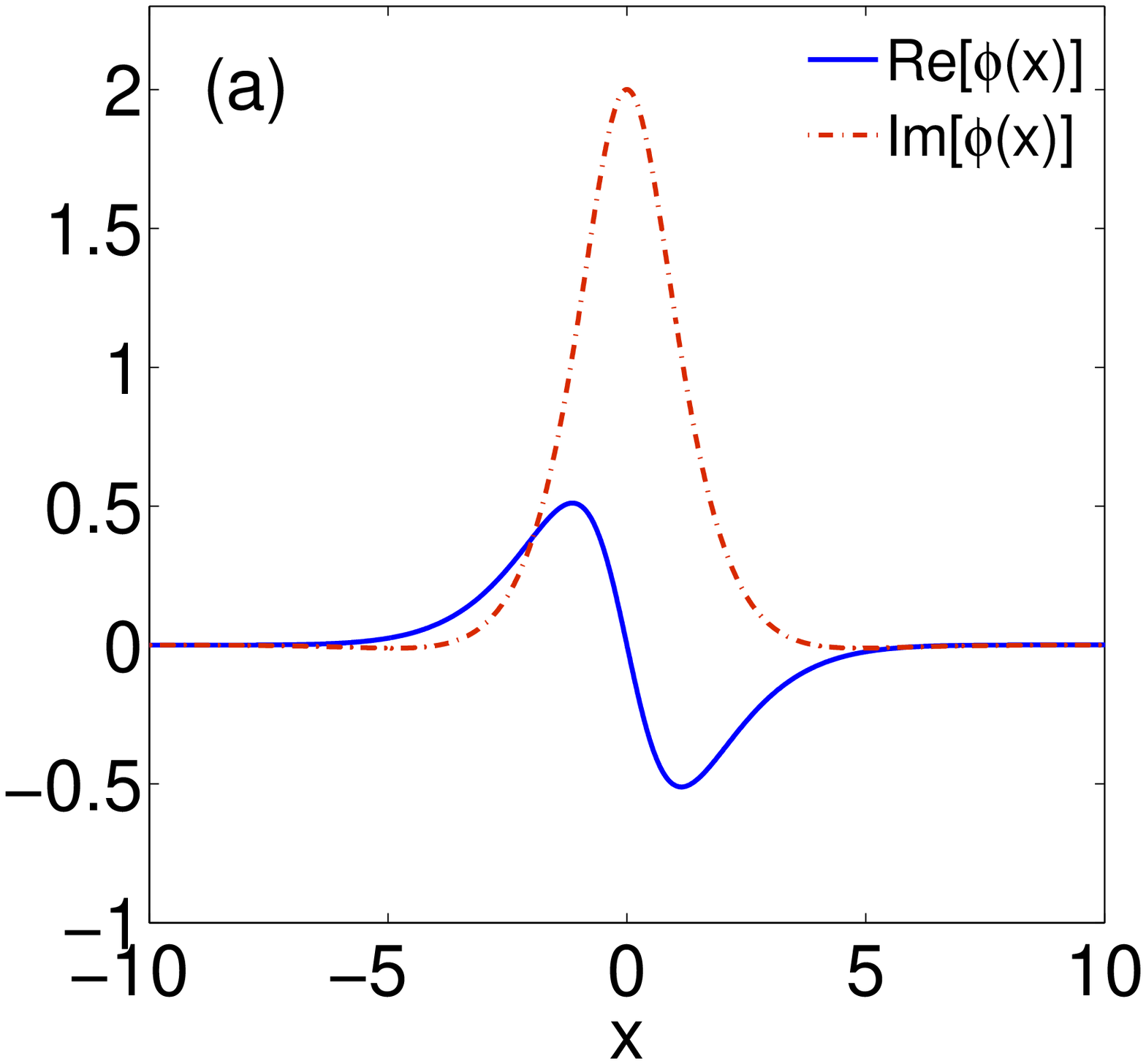} ~ ~~~~~~\includegraphics[width=5 cm, height=5 cm]{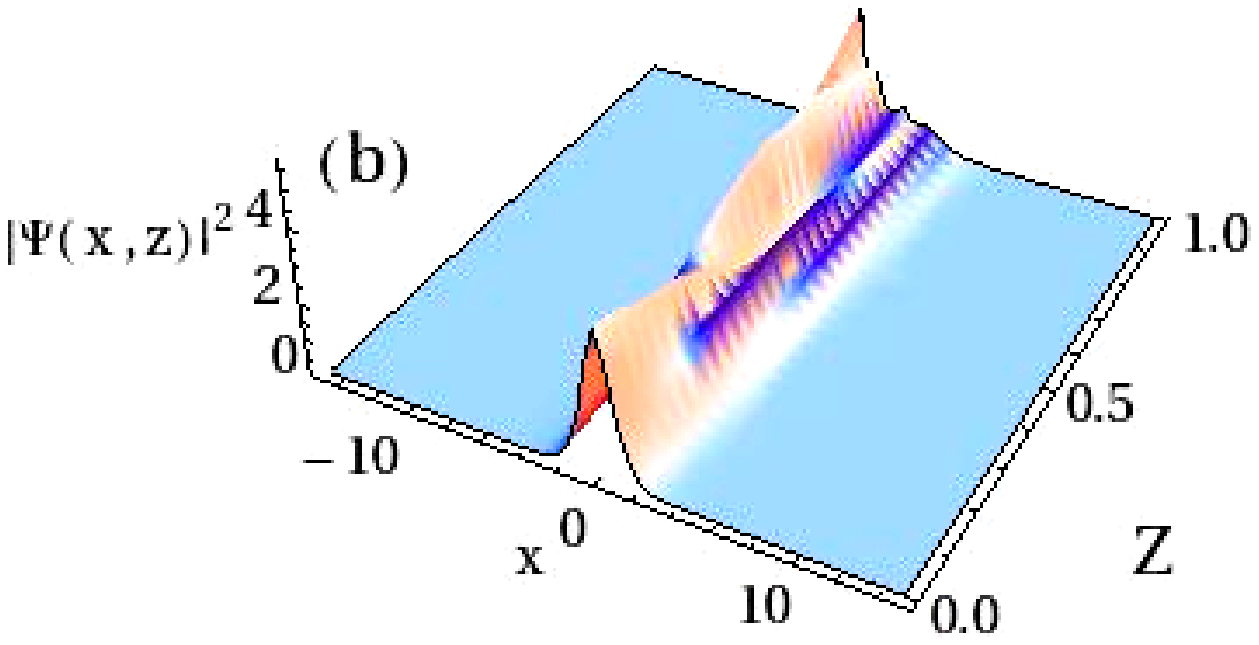} ~
~~~~~~~\includegraphics[width=5 cm,height=4.5 cm]{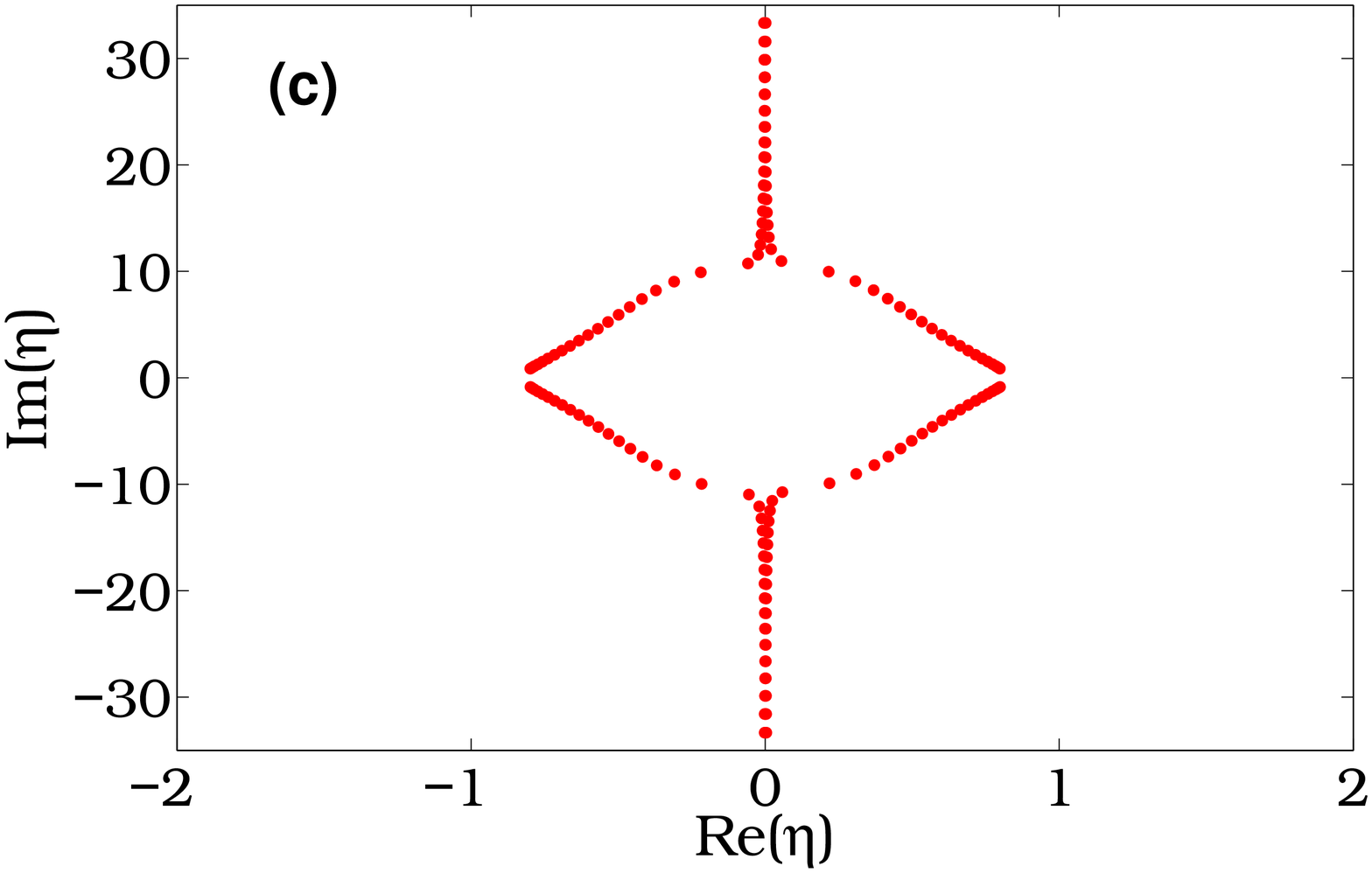}
\caption{(Color online) (a) Plots of the real (solid blue curve) and imaginary (dotted-dashed red curve) parts of the localized modes $\phi(x)$
in self-defocusing media; (b) Unstable intensity evolution $|\Psi(x,z)|^2$ corresponding to the figure \ref{f3}a; (c) Plot
of the corresponding stability
spectra obtained numerically. In all these cases we have considered potential parameters $a=1, b=.4$, $\lambda =.84$ and $\sigma=-1$.}\label{f3}
\end{figure}

Optical beam propagation in nonlinear self-defocusing Kerr medium is governed by the equation (\ref{e1}) for $\sigma=-1$ and its corresponding stationary solutions
satisfy equation (\ref{e3}). Like the self focusing  case, here equation (\ref{e3}) admits an exact solution
$\phi(x) = \sqrt{-(a^2+a+2)} ~\sech x ~e^{i b x}$. Note that these non-linear modes in self-defocusing case are very similar
to those in the self-focusing case. Only the amplitude of the localized modes are different. Nevertheless, like the self-focusing
case here also the localized modes are unstable for all $a,b \in \Re$.
In figures \ref{f3}(a), and \ref{f3}(b), we have shown such
nonlinear modes and corresponding unstable intensity evolution for the parameter values $a=1$, $b=.4$, $\lambda =.84$ and $\sigma =-1$.
Numerical solution of the
eigenvalue
problem (\ref{e12}) has been plotted in figure \ref{f3}(c) which also implies that the the corresponding modes are linearly unstable.
\FloatBarrier

\section{Localized modes in two-dimensions}
Finally, we discuss the formation of nonlinear localized modes in two-dimensional Rosen-Morse potential. The two-dimensional
generalization of the equation (\ref{e1}), with the self-focusing nonlinearity, is given by \cite{Mu+08b}
\begin{equation}
 i \frac{\partial \Psi}{\partial z} +  \nabla^2 \Psi + [V(x,y) + i W(x,y)] \Psi + |\Psi|^2 \Psi = 0, \label{e10}
\end{equation}
where $\nabla^2 \equiv \frac{\partial^2 }{\partial x^2} + \frac{\partial^2 }{\partial y^2}$ is the two-dimensional Laplacian.
The two-dimensional complex Rosen-Morse potential well, which obey the $\mathcal{PT}$-symmetric requirements $V(-x,-y) = V(x,y)$ and $W(-x,-y) = - W(x,y)$,
can be considered as
\begin{equation}
 \begin{array}{ll}
  V(x,y) = 2 (\sech^2 x + \sech^2 y) - (a^2+a+2) \sech^2 x ~ \sech^2 y\\
  W(x,y) =  4 b (\tanh x + \tanh y).
 \end{array}\label{e11}
\end{equation}
The stationary solutions of the equation (\ref{e10}) can be assumed in the form
\begin{equation}
 \Psi(x,y,z) = \phi(x,y) ~ e^{i \lambda z + i \theta(x,y)}
\end{equation}
where $\phi(x,y)$ and the phase $\theta(x,y)$
satisfy the following two equations
\begin{equation}\begin{array}{ll}
 \nabla^2 \phi - |\nabla \theta|^2 \phi + V(x,y) \phi + \phi^3 = \lambda \phi, \\
 \phi\nabla^2 \theta + 2 \nabla \theta . \nabla \phi + W(x,y)\phi  = 0.
\end{array}\label{e22}
\end{equation}
respectively.
\begin{figure}
 \includegraphics[width=4.25 cm,height=5 cm]{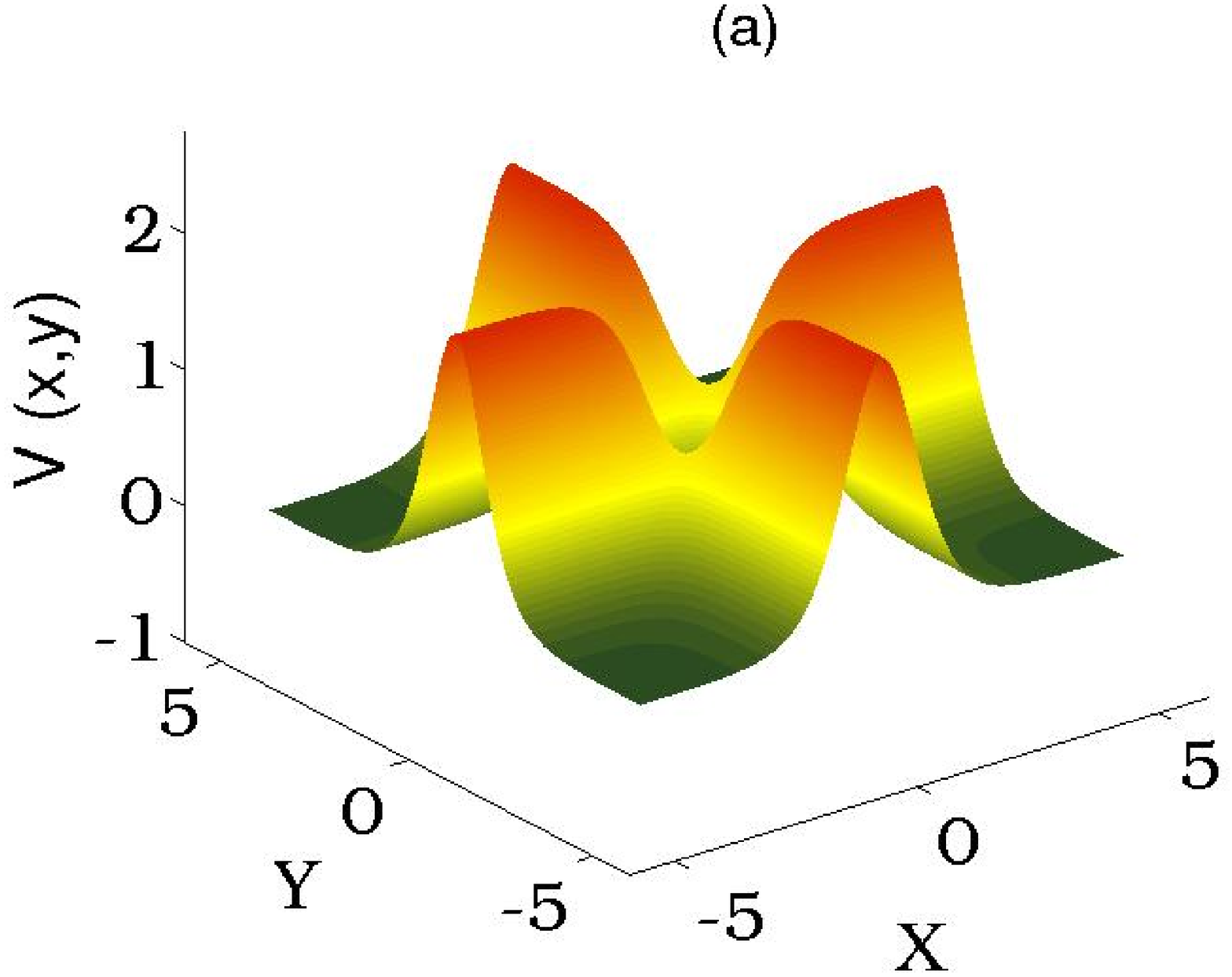} \includegraphics[width=4.25 cm,height=5 cm]{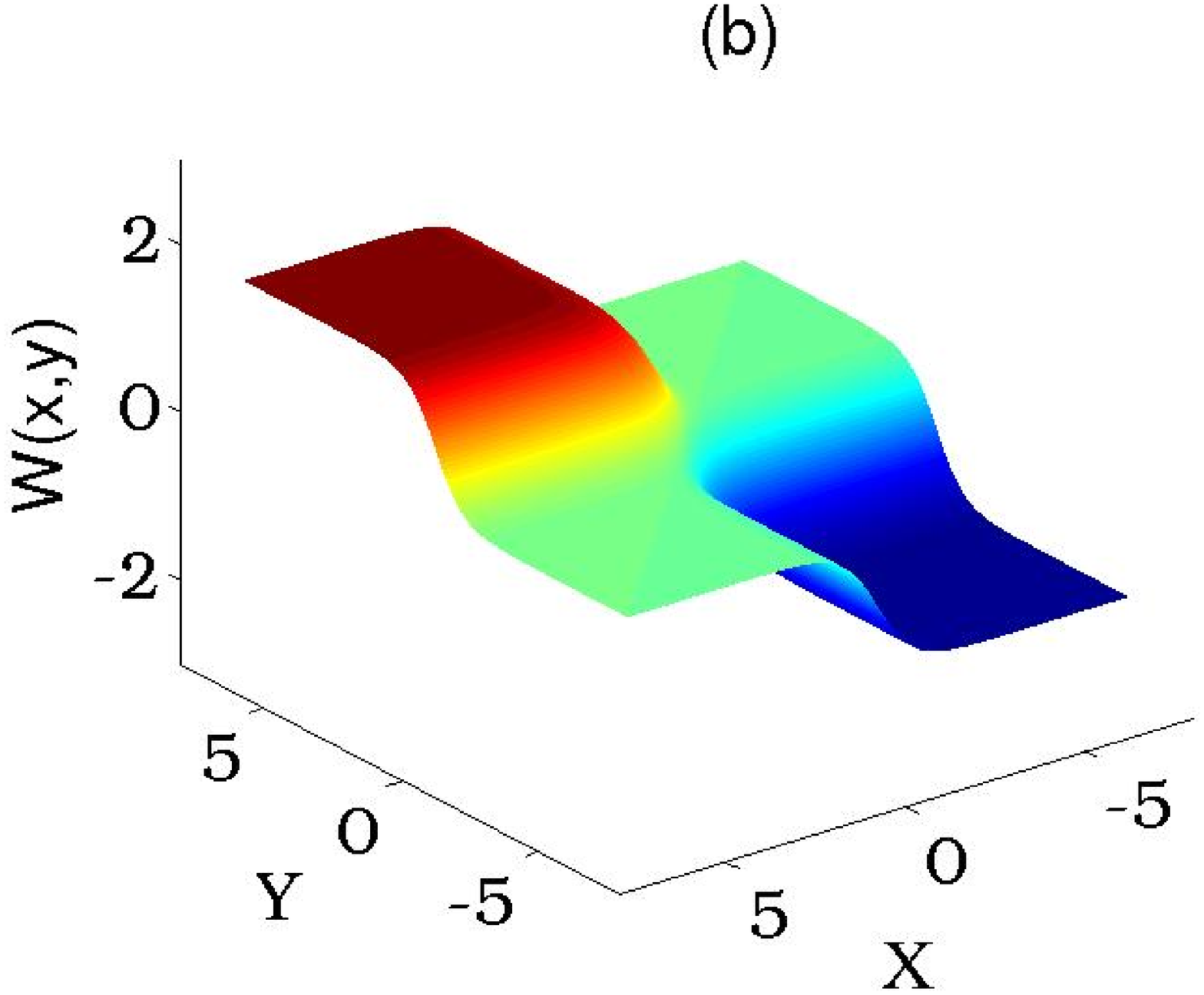}
   \includegraphics[width=4.5 cm,height=5 cm]{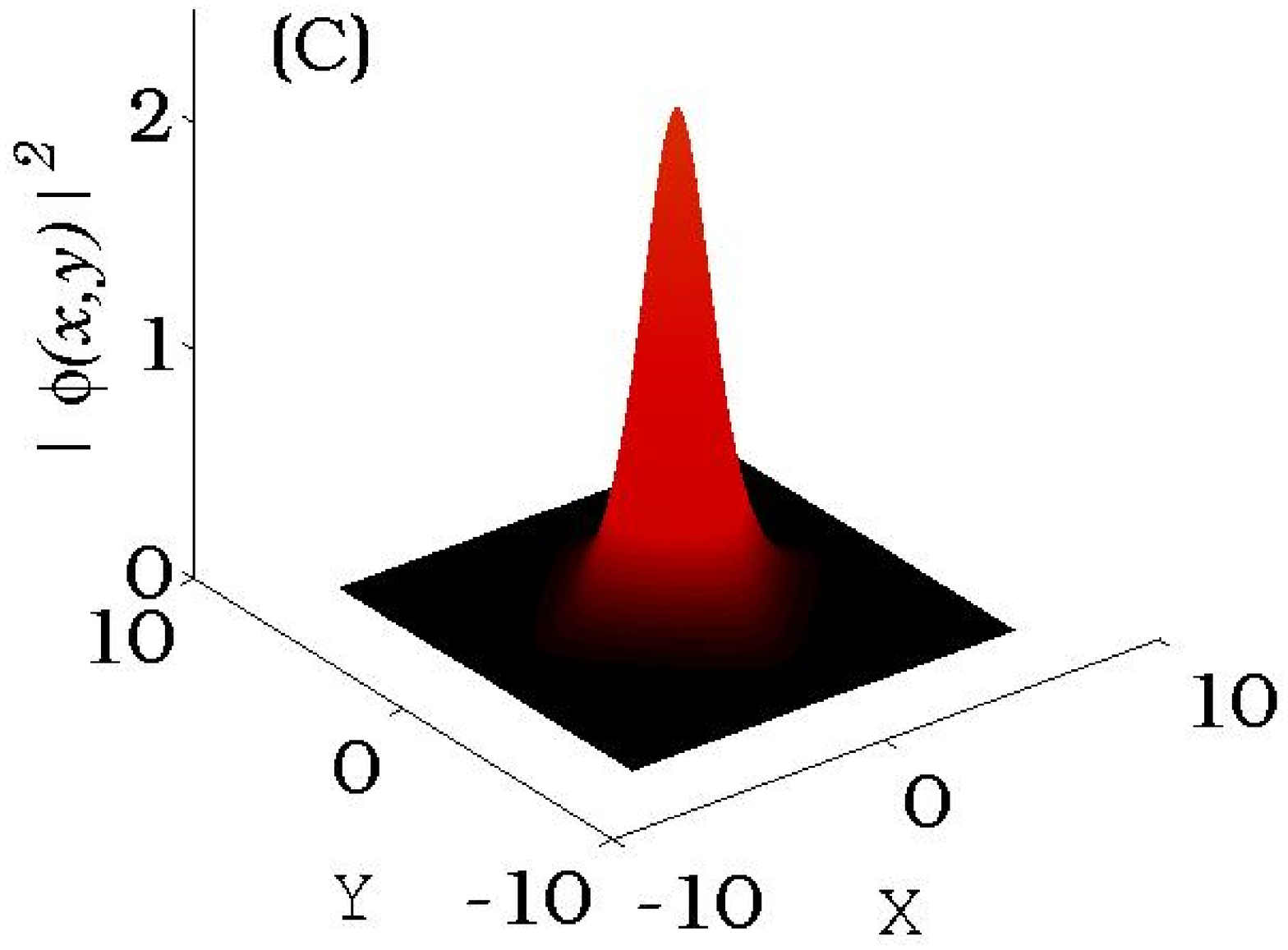} \includegraphics[width=4 cm,height=4.5 cm]{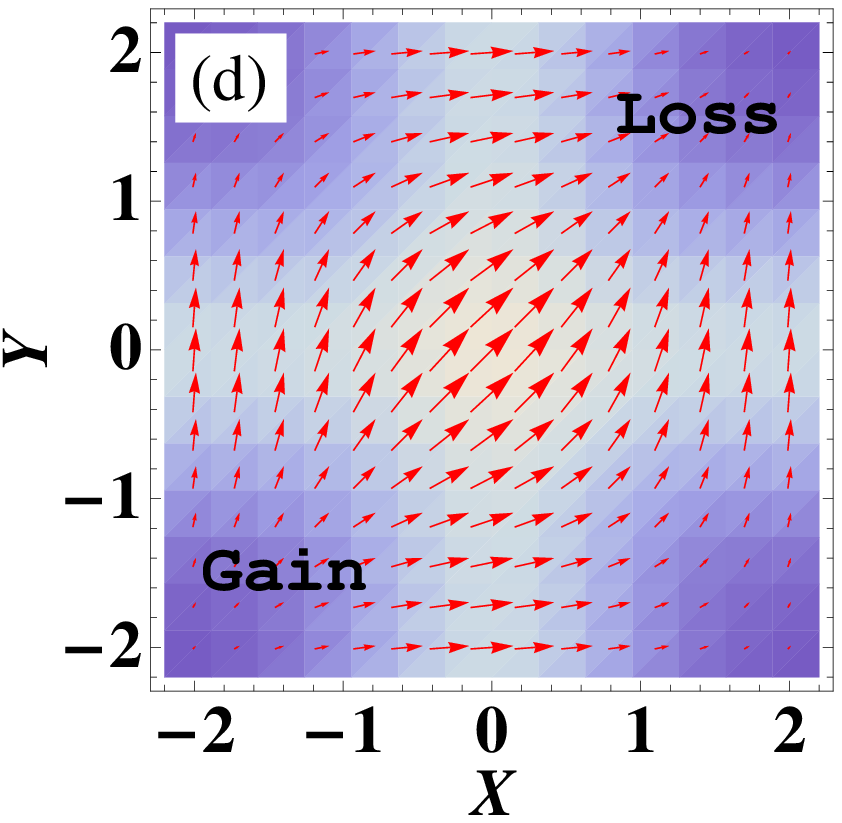}
 \caption{(Color online) (a) Real part of the 2D Rosen-Morse potential. (b) Imaginary part of the 2D Rosen-Morse Potential;
 (c) Plot of the 2D localized modes $|\phi(x,y)|^2$.
   (d) Transverse power flow vector $\vec{S}$ indicating the power flow from gain towards loss regions.
 In all these cases we have considered $a=1.25$ and $b=.5, \lambda = 1$ and $\sigma=1$.}\label{f5}
\end{figure}
\FloatBarrier
 A nonlinear solution to equation (\ref{e22}) that satisfies $\phi \rightarrow 0$ as $(x,y) \rightarrow \pm \infty$ is obtained as
\begin{equation}
 \phi(x,y) = \sqrt{a^2+a+2} ~ \sech x \sech y,
\end{equation}
with the phase $\theta(x,y)  =  b (x+y)$ and the propagation constant is given by $\lambda = 2- 4 b^2.$ Figures \ref{f5}(a),(b) show the real and imaginary parts of the 2D Rosen-Morse potential well. Two-dimensional soliton $|\phi(x,y)|^2$ is shown in figure $\ref{f5}(c)$. In all these
cases we have considered $a=1.75 $ and $\sigma=1$. To understand the internal structure of the two dimensional self-trapped modes, we calculate the
two-dimensional transverse power flow vector $\vec{S} = b(a^2+a+2) ~(\sech^2 x, \sech^2 y).$
In figure \ref{f5}(d), we have shown such 2D transverse power flow for $a=1.25, b=.5 \lambda=1, \sigma=1$, which
indicates energy exchange from gain towards loss regions.

\section{Summary}
 To summarize, we have investigated the existence and properties of nonlinear localized modes in a single $\mathcal{PT}$ waveguide cell characterized by the nonlinear Schr\"odinger
equation with complex Rosen-Morse potential well.
The closed form expressions for the localized modes in such one- and two-dimensional self-focusing and
self-defocusing Kerr nonlinear media are obtained.  The transverse power flow density is shown to remain
 positive for some parameter
values which indicates that power flow is in a single direction, mainly from gain towards loss regions.
 However, linear stability analysis of the one-dimensional
solitons reveals that these solitons are unstable over the whole
range of the potential parameter in spite of the fact that corresponding linear Schr\"odinger eigenvalue problem possesses unbroken
$\mathcal{PT}$-symmetry. The main reason behind such instability is that unlike the real part, the imaginary part of the complex Rosen-Morse potential well does not
vanish asymptotically. Therefore
any small fluctuation in the field intensity is amplified (or absorbed) which leads to the instability.
 The results presented here definitely encourage one to search for the stable localized modes
(if any) of the nonlinear Schr\"odinger equation with the $\mathcal{PT}$-symmetric Rosen-Morse potential well
in the presence of higher-order/competing or
other nonlinearities.

\section*{Acknowledgement} One of the authors (BM) thanks Dr. Barnana Roy for helpful discussions.


\begin{thebibliography}{101}
\bibitem{Re+12} A. Regensburger, C. Bersch, M. Miri, G. Onishchukov, D.N. Christodoulides,and U. Peschel, Nature {\bf 488}, 167 (2012).
\bibitem{Ru+10} C.E. Rutter, K.G. Makris, R.El-Ganainy, D.N. Christodoulides, M. Segev and D. Kip, Nature Physics {\bf 6}, 192 (2010).
\bibitem{Ko10} T. Kottos, Nature Physics {\bf 6}, 166 (2010).
\bibitem{Gu+09} A. Guo et al., Phys. Rev. Lett. {\bf 103}, 093902 (2009).
\bibitem{Ma08} K.G. Makris, R.El-Ganainy, and D.N. Christodoulides and Z.H. Musslimani, Phys. Rev. Lett. {\bf 100}, 103904 (2008).
\bibitem{Mu+08a} Z.H. Musslimani, K.G. Makris, R.El-Ganainy and D.N. Christodoulides, Phys. Rev. Lett. {\bf 100}, 030402 (2008).
\bibitem{Be08} M.V. Berry, J. Phys. A {\bf 41}, 244007 (2008).
\bibitem{Lo09} S. Longhi, Phys. Rev. Lett. {\bf 103}, 123601 (2009).
\bibitem{Ka+10} Y.V. Kartashov et al., Opt. Lett. {\bf 35}, 1638 (2010).
\bibitem{Ga07} R.El-Ganainy, K.G. Makris, D.N. Christodoulides, Z.H. Musslimani, Opt. Lett. {\bf 32}, 2632 (2007).
\bibitem{We+10} C.T. West, T. Kottos and T. Prosen, Phys. Rev. Lett. {\bf 104}, 054102 (2010).
\bibitem{LCV11} S. Longhi, Phys. Rev. B {\bf 80}, 235102 (2009).
\bibitem{Be+10} O. Bendix, R. Fleischmann, T. Kottos, and B. Shapiro, Phys. Rev. Lett. {\bf 103}, 030402 (2009).
\bibitem{KGM08} S. Klaiman, U. Gunther, and N. Moiseyev, Phys. Rev. Lett. {\bf 101}, 080402 (2008).
\bibitem{BB98} C.M. Bender and S. Boettcher, Phys. Rev. Lett. {\bf 80}, 5243 (1998).
\bibitem{Be07} C.M. Bender, Rept. Prog. Phys. {\bf 70}, 947 (2007).
\bibitem{Li+11} Z. Lin et al., Phys. Rev. Lett. {\bf 106}, 213901 (2011).
\bibitem{Lo11} S. Longhi, J. Phys. A {\bf 44}, 485302 (2011).
\bibitem{Mo13} A. Mostafazadeh, Phys. Rev. A 87, 012103 (2013).
\bibitem{Ma+10} K.G. Makris, R. El-Ganainy,D.N. Christodoulides, and Z.H. Musslimani, Phys. Rev. A {\it 81}, 063807 (2010).
\bibitem{MRR10} B. Midya, B. Roy and R. Roychoudhury, Phys. Lett. A {\bf 374}, 2605 (2010).
\bibitem{Mu+08b} Z.H. Musslimani et al., J. Phys. A {\bf 41}, 244019, (2008).
\bibitem{HH12} S. Hu and W. Hu, J. Phys. B {\bf 45}, 225401 (2012).
\bibitem{Sh+11} Z. Shi, X. Jiang, X. Zhu and H. Li, Phys. Rev. A {\bf 84}, 053855 (2011).
\bibitem{Zn00} M. Znojil, J. Phys. A {\bf 33}, L61 (2000).
\bibitem{LM09} G. Levai and E. Magyari, J. Phys. A {\bf 42}, 195302 (2009).
\bibitem{NGY12} S. Nixon, L. Ge, and J. Yang, Phys. Rev. A {\bf 85}, 023822 (2012).
\bibitem{ZKK11} D.A. Zezyulin, Y.V. Kartashov, V.V. Konotop, EuroPhys. Lett {\bf 96}, 64003 (2011).
\bibitem{Ya08} J. Yang, J. Comp. Phys. {\bf 227}, 6862 (2008).
\bibitem{Ya10} J. Yang, Nonlinear Waves in Integrable and Non-integrable Systems (SIAM, Philadelphia, 2010).
\bibitem{KMB12} A. Khare, S.M. Al-Marzoug and H. Bahlouli, Phys. Lett. A 376, 2880 (2012).
\bibitem{Fa+12} F.K. Abdullaev, Y.V. Kartashov, V.V. Konotop, and D.A. Zezyulin, Phys. Rev. A {\bf 83}, 041805(R) (2011).
\bibitem{Hu+11} S. Hu, X. Ma, D. Lu, Z. Yang, Y. Zheng, and Wei Hu, Phys. Rev A {\bf 84}, 043818 (2011).
\bibitem{Ac+12} V. Achilleos, P.G. Kevrekidis, D.J. Frantzeskakis and R.Carretero-Gonzalez, Phys. Rev. A {\bf 86} 013808 (2012).
\bibitem{ZK12} D.A. Zezyulin and V.V. Konotop, Phys. Rev. A {\bf 85}, 043840 (2012).
\bibitem{TTA12} E. N. Tsoya, S. Tadjimuratova, and F. Kh. Abdullaev, Optics Comm. {\bf 285} 3441 (2012).
\bibitem{Al+12} N.V. Alexeeva, I.V. Barashenkov, A.A. Sukhorukov, and Y.S. Kivshar, Phys. Rev. A {\bf 85}, 063837 (2012).
\bibitem{DM11} R. Driben, B.A. Malomed, Opt. Lett., {\bf 36}, 4323 (2011).
\bibitem{Zh+11} X. Zhu, H. Wang, L.X. Zheng, H.Li, and Y.J. He, Opt. Lett {\bf 36}, 2680 (2011).
\bibitem{WW11} H. Wang and J. Wang, Opt. Express {\bf 19}, 4030 (2011).
\bibitem{Li+11d} H. Li, Z. Shi, X. Jiang, and X. Zhu , Opt. Lett. {\bf 36}, 3290 (2011)


\end{thebibliography}
\end{document}